\documentclass[runningheads]{llncs}
\usepackage{cite}
\usepackage{hyperref}
\usepackage{pdflscape}
\usepackage{lscape}
\usepackage{amsmath,amssymb,amsfonts}
\usepackage{multicol, blindtext}
\usepackage{algorithmic}
\usepackage{graphicx}
\usepackage{float}
\usepackage{blindtext, rotating}
\usepackage{multirow}
\usepackage{subfig}
\usepackage{textcomp}
\usepackage{xcolor}
\usepackage{fancyhdr}
\usepackage{verbatim}
\usepackage{atbegshi}
\usepackage[export]{adjustbox}
\usepackage{longtable,rotating}
\usepackage{threeparttable}
\usepackage[linesnumbered,ruled,vlined]{algorithm2e}
\SetKwInput{KwInput}{Input}                
\SetKwInput{KwOutput}{Output}

\begin{document}
\title{Kryptonite: An Adversarial Attack using Regional Focus}

\titlerunning{Kryptonite Adversarial Attack}

\author{Yogesh Kulkarni, Krisha Bhambani}

\authorrunning{Yogesh et al.}
\institute{Pune Institute of Computer Technology \\ Dept. of Computer Engineering \\
\email{\{yogeshpict,krisha.bhambani\}@gmail.com}}
\maketitle              
\let\thefootnote\relax\footnotetext{Paper Accepted and Presented at ACNS'21 (Workshops), \href{https://doi.org/10.1007/978-3-030-81645-2_26}{\color{blue}{DoI}}}
\begin{abstract}
With the Rise of Adversarial Machine Learning and increasingly robust adversarial attacks, the security of applications utilizing the power of Machine Learning has been questioned. Over the past few years, applications of Deep Learning using Deep Neural Networks(DNN) in several fields including Medical Diagnosis, Security Systems, Virtual Assistants, etc. have become extremely commonplace, and hence become more exposed and susceptible to attack. In this paper, we present a novel study analyzing the weaknesses in the security of deep learning systems. We propose 'Kryptonite', an adversarial attack on images. We explicitly extract the Region of Interest (RoI) for the images and use it to add imperceptible adversarial perturbations to images to fool the DNN. We test our attack on several DNN's and compare our results with state of the art adversarial attacks like Fast Gradient Sign Method (FGSM), DeepFool (DF), Momentum Iterative Fast Gradient Sign Method (MIFGSM), and Projected Gradient Descent (PGD). The results obtained by us cause a maximum drop in network accuracy while yielding minimum possible perturbation and in considerably less amount of time per sample. We thoroughly evaluate our attack against three adversarial defence techniques and the promising results showcase the efficacy of our attack.

\end{abstract}

\section{Introduction}
Significant progress in the field of artificial intelligence has caused its use to be almost ubiquitous. From security and safety systems to autonomous cars and health care systems, incredible strides have been made to create efficient neural nets, however, they remain constantly vulnerable to adversarial attacks. Even though certain object detectors, classifiers, etc., have reached near human accuracy, it has been found that they can be easily fooled by small, almost unnoticeable modifications in images. As our reliance on artificial intelligence systems increases, the possible impact of their failure also increases tremendously. We need to ensure that these networks work in the most proper manner. New and more advanced attacks, where imperceptible perturbations are being added to disrupt the working of these networks are being made, against which existing defences are rendered useless. Hence, analysing every possible weakness in networks that make them susceptible to attack, is the need of the hour. 

Let us consider image specific tasks like image classification, object detection, etc. To accomplish these tasks, neural networks try to identify features characteristic to particular classes in the training set to best determine the result. An ideal adversarial attack is of course one that causes effective misclassification with minimum distortion. 

When a human attempts to classify objects in an image, their attention is inevitably drawn to the main region of interest, the object itself \cite{b1}. It stands to reason, that modifying this object may cause ambiguity in analysing the image. We use this same reasoning in the adversarial attack we present here.

In this paper, we emphasize more on medical datasets. Most medical datasets including MRIs, X-rays, etc. all contain a very specific, well-defined region of interest that can be analysed to detect and classify ailments. It has also been found that medical datasets are more vulnerable to adversarial attacks \cite{b2}. The inspiration for this attack is possible unexplored vulnerability of perturbations in a region of interest, since most adversarial attacks are fairly antagonistic to it. Medical datasets were hence chosen for experimentation of the attack for two reasons. The first reason for doing so is to demonstrate threats on a real life application of deep learning. The second reason is that  because the region of interest here can be easily extracted, monitored, and modified.

In this attack, we aim to monitor and encourage changes especially in the region of interest of the image in order to constrain the noise as much as possible to a particular area of the image. In order to prevent excessive changes in pixel values in a specific area, the perturbations are constrained. In most cases, this method effectively manages to fool the classifier with the addition of minimum noise. This attack hides in plain sight, and infected pixels cannot be easily identified, unlike other constricted attacks that use adversarial patches, which make it obvious to viewers that an image is adulterated. We also find that attacking this region of interest mitigates the effectiveness of state of the art defences, as compared to their effect on other attacks. 

To summarise the contribution to research presented in this paper:
\begin{itemize}
\item We propose a highly efficient and accurate, three-step RoI extraction algorithm built upon Otsu's method of image thresholding, image dilation using predefined kernel and finally extract RoI using the contour of lesion/tumor found using Topological Structural Analysis.

\item We propose Kryptonite, a white-box, non-targeted adversarial attack, that exploits a proposed area of weakness.

\item We compare our proposed attack with the existing state of the art attacks based on Receiver Operating Characteristic Curve (ROC), and perturbation size.

\item We analyse the comparison of the impact of various state of the art defences on the proposed attack with other attacks.

\item We aim to show the advantages of localising an attack in an image.
\end{itemize} 

The remainder of the paper is structured as follows: Section \ref{Related_Work} talks about related work. Section \ref{model} elaborates on the dataset and network architecture used. Section \ref{attack_overview} provides an overview of our proposed attack. In Section \ref{proposed} the proposed methodology for RoI extraction and Kryptonite is discussed. Section \ref{results} highlights our obtained results, Section \ref{limitations} highlights the limitations of our study and Section \ref{conclusion} concludes the paper with the scope of further research and improvement.

\section{Related Work} \label{Related_Work}

\subsection{Adversarial Attacks}
\subsubsection{FGSM} 

The Fast Gradient Sign Method proposed by \cite {b3} uses a method where a given loss function J(x, y), which is almost always cross-entropy loss that the target network is trained on, is maximised, and the sign of the gradient obtained is used to control the added perturbation of adjustable magnitude in $L_\infty$ norm.
\begin{equation}
        x* = x + \epsilon . sign(\nabla_x J(x, y))
\end{equation} 

\subsubsection{I- FGSM} 
The proposed Iterative Fast Gradient Sign Method proposed by \cite{b4} is a simple, but a greatly effective improvement over the simple FGSM attack. It uses the observation that  for any input image x, the gradient of the adopted loss function is continuously changing. Hence in every iteration, a perturbation is generated that is optimal for the current gradient obtained. Here, a clip function is used to control the size of the final perturbation and restrict it to the original constraint $\epsilon$.
\begin{equation}
 x*_{t+1} = Clip_{x,\epsilon} \{x*_t + \alpha . sign(\nabla_x J(x*_t, y))\} 
\end{equation}
Projected Gradient Descent proposed by \cite{b5} is a variation of FGSM where the constraint $\alpha.T = \epsilon$ does not exist. Instead, to constrain perturbations, the adversarial samples are "projected" to their benign counterparts. Images in PGD are updated as follows:
 \begin{equation}
x*_{t+1} = Proj\{x*_t + \alpha . sign(\nabla_x J(x*_t, y))\}
 \end{equation} 
PGD is more powerful than FGSM but it is slower than FGSM as it calculates gradients for numerous iterations.

\subsubsection{MI- FGSM} 
Momentum Iterative Attack proposed by \cite{b6} is a version of I-FGSM that uses the technique of memorization of previous gradients to optimise the iterative process and find the most effective perturbation. This helps to stabilise update directions and escape local minima that may yield poor, less-than-ideal perturbations.

\begin{equation}
g_{t+1} = \mu * g_t + \frac{\nabla_x J(x, y))}{||\nabla_x J(x,y))||_1}
\end{equation}
\begin{equation}
 x*_{t+1} = Clip_{x,\epsilon} \{x*_t + \alpha . sign(g_{t+1})\} 
\end{equation}

\subsubsection{DeepFool} 
DeepFool proposed by \cite{b7} is an adversarial attack based on the assumption that the neural networks are linear, and that the various classes are essentially separated by a hyperplane. Here, imperceptible perturbations are added to take a step forward to push the sample to be classified over the decision boundary. Since most neural networks are not linear, this is done iteratively until the adversarial example is optimally constructed. In this paper, we use the $L_\infty$ version of the attack.

\subsection{Adversarial Attacks and Defences on Medical Imaging}
The intensive research in \cite{b2} showed adversarial attacks on medical images are \textit{easy} to detect, we \textit{improved} upon their research by exploiting image regions and producing more erroneous adversarial features which easily fool DNN's and anomaly detectors. An Ensemble of multiple Convolutional Neural Network's (CNN) and inclusion of adversarial images while training was proposed in \cite{b8} for mitigating adversarial attacks against \textit{simpler} FGSM and One - Pixel attacks but the inclusion of adversarial images while training may \textit{not} provide resilience as shown later in table \ref{defence_table}. DL systems in production consist of complex data pipelines, therefore, the addition of new component as proposed in \cite{b9} for detection of adversarial attacks on medical images would be \textit{arduous}, also as proved by Carlini and Wagner in \cite{b10} the proposed unsupervised statistical anomaly detection technique can easily be evaded when an adversary targets a specific defence which is a white box adversary in this case. 
A novel adversarial bias field attack was proposed by \cite{b11} for chest X-ray classification systems, which generated more realistic adversarial samples by adding smooth perturbations instead of noises but the attack success rate was \textit{less} than noise-based adversarial attacks.

\subsection{Analysis of Related Attacks}
There have been very few attacks focused on modifying a constrained area of an image. One such attack is mentioned in \cite{b12}, which essentially uses gradient information to adjust the trust region, within a continuously adaptive radius. With this, the extent of the added noise in the image is restricted, however, the \textit{JumpReLU} defence proposed by \cite{b13} has provided resilience to this attack as well. Another attack in this category is the localized BIM attack proposed by \cite{b14}, wherein the final perturbations were hard to detect yet \textit{efficacy of the proposed attack is questionable} as the attack has more emphasis on minimal human perceptibility and the paper lacks evaluation with existing adversarial attacks in terms of drop in network accuracy.
We hence believe our attack is a novel, efficient and effective addition to this limited class of families.

\section{Dataset and Model Architecture} \label{model}
We have presented results for 2 datasets and 4 neural nets in this paper. 
The first dataset is the Melanoma dataset, whereas the other dataset is the MRI dataset.
 
The first medical dataset we have used is the 2020 ISIC Challenge Dataset \cite{b15} containing 33,126 dermoscopic training images having unique benign and malignant lesions from over 2,000 patients. We split the data into training, validation and testing sets having 20,000, 6,576 and 6,550 samples respectively. We generated adversarial samples from the test set. Data augmentations the images went through include cutout, hue saturation, addition of Gaussian noise, motion or median blur, optical or grid distortion, as well as rotation and flipping. We have used EfficentNet-B5 \cite{b16} and ResNeXt-50 \cite{b17} as these architectures are extremely complex, and often yield state of the art results for biomedical imaging datasets as seen in table \ref{dermatology_auc_table}.

\begin{table}[t]
\caption{CNN Architecture.}
\centering 
\begin{tabular}{| c | c |}
\hline
\label{CNN_architecture}
Layers & Parameters \\
\hline
Input & 126x126x1 \\
Conv1 + ReLU & 50x3x3, pad = same, stride = 1 \\
Conv2 + ReLU & 75x3x3, pad = same, stride = 1 \\
Max Pool 1 & 2x2, stride 2, pad = 0 \\
Dropout & 0.25 \\
Conv3 + ReLU & 125x3x3, pad = same, stride = 1 \\
Max Pool 2 & 2x2, stride 2, pad = 0 \\
Dropout & 0.25 \\
FC1 + ReLU & 500 \\
Dropout & 0.4 \\
FC2 + ReLU & 250 \\
Dropout & 0.3 \\
FC3 + Sigmoid & 1 \\
\hline
Total Parameters & 60, 307, 326 \\
\hline
\end{tabular}
\end{table}

The second dataset we have used is the MRI dataset \cite{b18}. Here we aim to identify whether a given MRI of a brain has a tumor or not, and classify it respectively. The original training set contains 253 images of MRI scans out of which 155 have a tumour and 98 do not. We augment the dataset to generate 2065 images out of which 1200 are used for training the neural network, 365 are used for validation and 500 are used for testing. 
To classify images for this dataset, we have used the VGG16 \cite{b19} neural network, which gives near state of the art results. The second network we use to execute this task is a custom convolutional neural network. The architecture for the same is provided in table \ref{CNN_architecture}. This also yields fairly accurate classification results, as shown later in table \ref{radiology_auc_table}.

\section{Attack Overview} \label{attack_overview}
\subsection{Threat Model}
Today, several effective defences have been created to avoid attacks, however, attacks to circumvent these defences are being continuously created. For example, the CW attack \cite{b20} was able to render defensive distillation almost useless.

It is important to analyse the vulnerability of any image, which is what we aim to do with Kryptonite. 
A vulnerability we exploit through the attack Kryptonite is one that has been largely ignored by several attacks, and hence defences: the region of interest. For this attack we assume that the attacker \textit{has access to every aspect} of the architecture of the network and its parameters, and hence assume this attack is \textit{white box}. Kryptonite launches its attack by focusing mainly on the region of interest. In most cases, this method effectively manages to fool the classifier with the addition of minimum noise. 

Kryptonite belongs to a class of iterative gradient based attacks, and it aims to show the ways in which the effectiveness of an attack can be improved by localising it. We have hence compared Kryptonite to such attacks in terms of drop in accuracy, effectiveness, efficiency, and resilience to state of the art defences, and found that focusing on a region of interest indeed improves performance on all fronts. 

\subsection{Metrics Used}
The distance metric we have chosen to use to evaluate and perform this attack is the $L_p$ norm metric. 
$L_p$ distance is expressed as:
\begin{equation}
     ||x - x’||_p = (\sum_{i=1}^{n} |x - x’|)^{1/p}
\end{equation}
 The metric is used to limit the maximum change in pixel values, and evaluate the size of the perturbation by measuring the Euclidean distance between the original and perturbed image.
 
\section{Proposed Methodology} \label{proposed}
\subsection{Extracting Region of Interest}
The Region of Interest extractor consists of the following modules:
\subsubsection{OTSU Thresholding method} We use the methodology given in \cite{b21} to binarize the image based on pixel intensities and separate the pixels into classes, foreground and background. We minimize the weighted within-class variance in search of optimal threshold which is given below:
\begin{equation}
  \sigma_{w}^{2}(t)=\omega_{0}(t) \sigma_{0}^{2}(t)+\omega_{1}(t) \sigma_{1}^{2}(t)
\end{equation}
In the above equation \( \omega_0\) and \( \omega_1\) are probabilities of two classes which are separated by threshold \( t\), and \( \sigma_0^2\) and \( \sigma_1^2\) are the variances for these two classes. For L bins of the histogram we compute the class probability \( \omega_{0,1}(t)\) as follows:
\begin{gather}
    \omega_{0}(t)=\sum_{i=0}^{t-1} p(i)\\
    \omega_{1}(t)=\sum_{i=t}^{L-1} p(i)
    \end{gather}
We minimize the intra-class variance as follows:
\begin{equation}
    \begin{split}
      \sigma_{b}^{2}(t) &=\sigma^{2}-\sigma_{w}^{2}(t)=\omega_{0}\left(\mu_{0}-\mu_{T}\right)^{2}+\omega_{1}\left(\mu_{1}-\mu_{T}\right)^{2}\\   
      & = \omega_{0}(t) \omega_{1}(t)\left[\mu_{0}(t)-\mu_{1}(t)\right]^{2}
    \end{split}
\end{equation}

\begin{figure}[h]
\centering
\subfloat{\includegraphics[width = .16\linewidth]{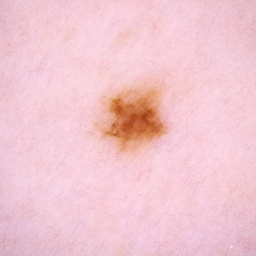}} 
\subfloat{\includegraphics[width = .16\linewidth]{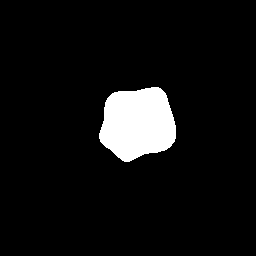}}
\subfloat{\includegraphics[width = .16\linewidth]{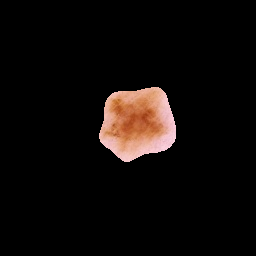}}
\hspace{1mm}
\subfloat{\includegraphics[width = .16\linewidth]{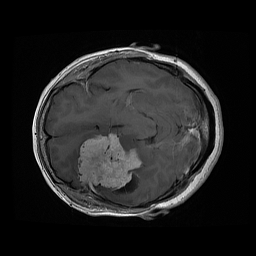}} 
\subfloat{\includegraphics[width = .16\linewidth]{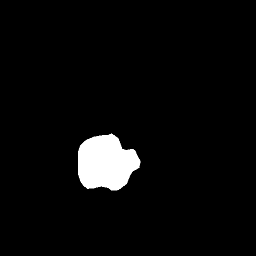}}
\subfloat{\includegraphics[width = .16\linewidth]{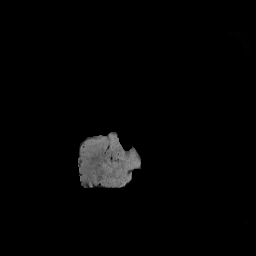}}
\caption{Process for generation of the RoI on original image, application of OTSU Thresholding and dilation, and finally, derived RoI}
\label{process_fig}
\end{figure}

The above expression is expressed in terms of class probabilities \(\omega\) and the class means \(\mu\), and the class means are defined as follows:
\begin{gather}
    \mu_{0}(t)=\frac{\sum_{i=0}^{t-1} i p(i)}{\omega_{0}(t)}\\
    \mu_{1}(t)=\frac{\sum_{i=t}^{L-1} i p(i)}{\omega_{1}(t)}\\
    \mu_{T}=\sum_{i=0}^{L-1} i p(i)
    \end{gather}
We compute the class probabilities and class means iteratively. Algorithm \ref{alg1} explains briefly the procedure to minimize the weighted within-class variance.
\begin{algorithm}[t]
\SetAlgoLined
\KwInput{Grayscale image}
  \KwOutput{Intensity Threshold}
  For each intensity level compute histogram and intensity level probabilities.\\
  Initialize \(\omega_i(0)\) and \(\mu_i(0)\).\\
 \For{threshold t = 1, .......upto max(intensity)}{
  update \(\omega_i\) and \(\mu_i\)\;
  Compute \(\sigma_b^2(t)\)\;
 }
 return max(\(\sigma_b^2(t)\)) which corresponds to final threshold value.
 \caption{OTSU Thresholding}
 \label{alg1}
\end{algorithm}
As we can see from figure \ref{process_fig}, the RoI extraction algorithm accurately returns the RoI. The algorithm is simple yet effective and is highly accurate. Although, one can obtain decent results using complex segmentation models such as U - Net \cite{b22} by training neural networks on enormous datasets but our method is much simpler and efficient which leads to an stronger region constricted adversarial attack.

\subsubsection{Image Dilation}
To increase the object area and to accentuate the features we performed image dilation. To achieve the same we performed the following steps:
\begin{itemize}
\item We convolve the thresholded image obtained from algorithm \ref{alg1} with a kernel (matrix of odd size).
\item We define the center of the kernel as the anchor point.
\item We scan the kernel over the image and calculate the maximum value of the pixel overlapped by the kernel and replace the image pixel at the anchor point position with the maximum value. This would increase the white region in the image and the size of the foreground object.
\end{itemize}
\subsubsection{Topological Structural Analysis}
We have utilized the contour tracing algorithm given by Suzuki \cite{b23}. The algorithm defines hierarchical relationships among the borders and differentiates between the outer and the hole boundary. As this is an iterative algorithm it connects groups of 1-pixels that surround groups of 0-pixels. Then, a raster scan of the image is performed which locates all possible pixels for a border \cite{b24}, by detecting whether the pixel has value 1 and the neighboring pixel has value 0. Then a label is assigned to keep track of the border. Then for each new succeeding pixel, it is either added to an existing border or a new border with a new label. If two border segments connect then we reassign labels of  pixels to form one border. We return the contour obtained for skin lesion and the pixels inside it. For a 512x512x3 image from \cite{b15}, RoI extraction took \textit{1.99 milliseconds} whereas for 126x126x1 grayscale image from \cite{b18}, RoI extraction took \textit{0.88 milliseconds}.

Our proposed RoI extraction algorithm performs perfectly for images having distinct contour lines but noisy images or images having no contour lines surrounding the RoI, our algorithm returns \textit{arbitrary
}RoI.

\subsection{Kryptonite:}
Kryptonite comes under a class of momentum iterative gradient-based methods, and hence is compared to similar attacks in this paper. Kryptonite is essentially an adversarial attack proposed to improve adversarial attacks boosted by momentum even further, by monitoring the changes observed specifically in the region of interest. 
Kryptonite uses a region of interest extractor($\rho$) that specifically monitors these features to evaluate the progress(P) of the attack. The change of these features is used to optimise the momentum applied to simple Iterative Fast Gradient Sign Method.
Essentially, this network aims to demonstrate the increased susceptibility of the images, by monitoring the region of interest.
Momentum is a method used to optimise the efficiency of gradient descent and provide a certain acceleration to the algorithm to help it easily navigate through local minima, and other hurdles effectively. 

Kryptonite uses the region of interest($\rho$) method to evaluate the attack’s exact progress(P) as:

\begin{equation}
	 P =\|  \rho(x*_{t+1}) - \rho(x*_{t})  \|_2
\end{equation}
This progress in the absolute area of interest is used to determine the decay factor for the next iteration. The decay factor can be written as:
\begin{equation}
\mu_{t+1} = \frac{1}{ P } * \omega
\end{equation}
Where $\omega$ is the decay weight. This is a hyper parameter to be specified by the user according to the network requirements.
If this hyper parameter is too small, the network will be caused to act like Iterative Fast Gradient Sign Method. If the gradient is too large, the inertia of the attack would increase which would not allow any significant progress to be made. 

\begin{algorithm}[t]
\SetAlgoLined
\KwInput{Image x, Classifier f with loss function J, Region of Interest Extractor $\rho$, Size of perturbation $\epsilon$, Iterations T, Decay Weight $\omega$, Initial decay factor $\mu_0$}
\KwOutput{Perturbation x* clipped as \( \|x* - x\|_\infty \leqslant \epsilon\)} 
    Initialize $g_0= 0, x*_0=x $ and \(\alpha=\frac{\epsilon}{T}\). \\
    \For{threshold t = 0 to (T-1)}{
        Input $x*_t$ to f and retrieve \(\nabla_x J(x, y)\). \\
        Update $g_{t+1}$ by accumulating the velocity vector in the gradient direction:
        \[g_{t+1} = \mu_t * g_t + \frac{\nabla_x J(x, y))}{||\nabla_x J(x,y))||_1}\]\\
        Update $x*_{t+1}$ by applying sign gradient:
        \[x*_{t+1} = Clip_{x,\epsilon} \{x*_t + \alpha . sign(g_{t+1}) \}\]\\
        Calculate progress P:
        \[P = \| \rho(x*_{t+1}) - \rho(x*_{t}) \|_2\]\\
        Update decay factor:
        \[\mu_{t+1} = \frac{1}{P} * \omega\]\\
    }
    return $x*_{T-1}$
\caption{Kryptonite adversarial attack}
\label{alg2}
\end{algorithm}

The gradient is then updated using this newly obtained decay factor in a similar way as that used in MI-FGSM.
\begin{equation}
	g_{t+1} = \mu_t * g_t + \frac{\nabla_x J(x, y))}{||\nabla_x J(x,y))||_1}
\end{equation}

This updated gradient is used to assess the optimal perturbation to add to the original image.
\begin{equation}
x*_{t+1} =  x*_t + \alpha . sign(g_{t+1})
\end{equation}

\begin{figure}[t]
\centering
\begin{tabular}{cc}
  \includegraphics[width=60mm]{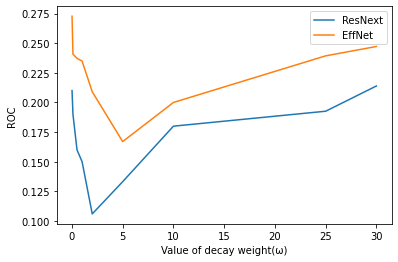} & 
   \includegraphics[width=60mm]{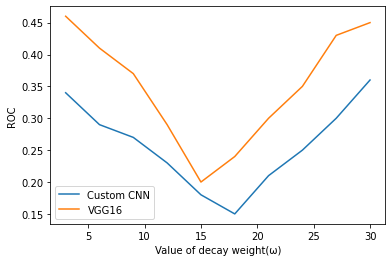} \\
    (a) Dermatology Dataset & (b) Radiology Dataset \\
\end{tabular}
\caption{Comparison of ROC score to variation in decay wt.}
\label{roc_vs_w_all}
\end{figure}

\begin{table*}[b]
\tiny
\caption{Optimal Hyper Parameters for all the attacks for the given two datasets.}
\label{hyper_parameter_table}
\centering
\begin{threeparttable}

\begin{tabular}{| c | c | c | c | c | c | c | c | c | c | c | }
\hline
\multirow{2}{*}{Melanoma} & \multicolumn{2}{|c|}{Epsilon} & \multicolumn{2}{|c|}{Iterations} 
& \multicolumn{2}{|c|}{Alpha} & 
 \multicolumn{2}{|c|}{Decay Factor} & \multicolumn{2}{|c|}{Overshoot}\\
\cline{2-11}
& EffNet & ResNext & EffNet & ResNext & EffNet & ResNext & EffNet & ResNext & EffNet & ResNext\\
\hline
FGSM & 0.08 & 0.07 & -- & -- & -- & -- & -- & -- & -- & --\\
\hline
DeepFool & -- & -- & 60 & 45 & -- & -- & -- & -- & 0.07 & 0.04\\
\hline
PGD & 0.04 & 0.03 & 16 & 12 & 0.04/16 & 0.03/12 & -- & -- & -- & --\\
\hline
MIFGSM & 0.03 & 0.01 & 10 & 7 & 0.04/10 & 0.02/7 & 1.5 & 1 & -- & --\\
\hline
KRYPTONITE & 0.01 & 0.007 & 15 & 5 & 0.01/15 & 0.007/5 & *0.5 & *1 & -- & --\\
\hline
\multirow{2}{*}{MRI} & \multicolumn{2}{|c|}{Epsilon} & \multicolumn{2}{|c|}{Iterations} 
& \multicolumn{2}{|c|}{Alpha} & 
 \multicolumn{2}{|c|}{Decay Factor} & \multicolumn{2}{|c|}{Overshoot}\\
\cline{2-11}
& CNN & VGG16 & CNN & VGG16 & CNN & VGG16 & CNN & VGG16 & CNN & VGG16\\
\hline
FGSM & 0.1 & 0.15 & -- & -- & -- & -- & -- & -- & -- & --\\
\hline
DeepFool & -- & -- & 50 & 65 & -- & -- & -- & -- & 0.06 & 0.08\\
\hline
PGD & 0.03 & 0.05 & 20 & 12 & 0.03/20 & 0.05/12 & -- & -- & -- & --\\
\hline
MIFGSM & 0.02 & 0.03 & 12 & 15 & 0.02/12 & 0.03/15 & 0.5 & 1 & -- & --\\
\hline
KRYPTONITE & 0.02 & 0.04 & 16 & 20 & 0.02/16 & 0.04/20 & *0.5 & *0.7 & -- & --\\
\hline
\end{tabular}
\begin{tablenotes}
      \tiny
      \item[*] Inital decay factor for Kryptonite
    \end{tablenotes}
\end{threeparttable}

\end{table*}

\begin{table}[t]
\caption{The table below shows the average of network accuracy after conducting the experiment five times on the Dermatology Dataset.}
\label{dermatology_auc_table}
\centering
\begin{threeparttable}
\begin{tabular}{| c | c | c | c | c | }

\hline 
\multirow{2}{*}{Samples} & \multicolumn{2}{|c|}{Effnet} & \multicolumn{2}{|c|}{ResNext} \\ [0.5ex] 
\cline{2-5}
& Accuracy & Size of Pert. & Accuracy & Size of Pert. \\
\hline
CLEAN & 0.881 & -- & 0.890 & --\\
\hline
FGSM & 0.442 & 5.2\%/5.9\% & 0.391 & 4.7\%/5.5\%\\
\hline
DeepFool & 0.419 & 1.5\%/2.3\% & 0.406 & 1.2\%/1.4\% \\
\hline
PGD & 0.273 & 2.3\%/2.8\% & 0.206 & 1.9\%/2.5\%\\
\hline
MIFGSM & 0.229 & 2.6\%/3.0\% & 0.187 & 2.4\%/2.9\% \\
\hline
KRYPTONITE & \textbf{0.155} & {1.7\%/2.2\%} & \textbf{0.114} & {1.5\%/1.7\%}\\ [1ex]
\hline
\end{tabular}
\begin{tablenotes}
     \tiny
     \item Size of perturbation is expressed as a percentage of added perturbation to image measured using the standard $L_2$ norm. The value of the left of the slash is average case percentage perturbation, and to the right is the worst case percentage perturbation.
\end{tablenotes}
\end{threeparttable}
\end{table}

\begin{table}[ht]
\caption{The table below shows the average of network accuracy after conducting the experiment six times on the Radiology Dataset.}
\label{radiology_auc_table}
\centering
\begin{threeparttable}

\begin{tabular}{| c | c | c | c | c | }
\hline 
\multirow{2}{*}{Samples} & \multicolumn{2}{|c|}{Custom CNN} & \multicolumn{2}{|c|}{VGG16} \\ [0.5ex]
\cline{2-5}
& Accuracy & Size of Pert. & Accuracy & Size of Pert. \\
\hline
CLEAN & 0.942 & -- & 0.966 & --\\
\hline
FGSM & 0.474 & 7.0\%/7.6\% & 0.490 & 7.3\%/7.8\%\\
\hline
DeepFool & 0.379 & 1.8\%/2.5\% & 0.399 & 2.1\%/2.9\% \\
\hline
PGD & 0.229 & 2.5\%/3.0\% & 0.293 & 2.6\%/3.1\%\\
\hline
MIFGSM & 0.161 & 2.8\%/3.4\% & 0.247 & 3.0\%/3.7\% \\
\hline
KRYPTONITE & \textbf{0.147} & {1.9\%/2.6\%} & \textbf{0.216} & {2.0\%/2.7\%}\\ [1ex]
\hline
\end{tabular}
 
\begin{tablenotes}
     \tiny
      \item Size of perturbation is expressed as a percentage of added perturbation to image measured using the standard $L_2$ norm. The value of the left of the slash is average case percentage perturbation, and to the right is the worst case percentage perturbation.
\end{tablenotes}
\end{threeparttable}

\end{table}

\section{Results} \label{results}
\subsection{Attack}
We perform all attacks in the white-box setting. We used the Cleverhans library \cite{b25} for the implementation of the attack algorithms. The optimal hyper parameters for all the attack algorithms are provided in Table \ref{hyper_parameter_table}. These parameters were chosen by monitoring network accuracy, perturbation size and actual perceptibility of perturbations to the naked eye. A small increment to optimal values of epsilon or increasing the number of iterations will lead to further drop in accuracy but increase perturbation size and also the final adversarial image may look like an image someone has tampered with. Furthermore, The optimal decay weight for kryptonite can be observed from figure \ref{roc_vs_w_all}. As can be observed from figure \ref{roc_vs_w_all}, this parameter makes a great difference to the reduction in ROC score, and the optimal value must be chosen for the best results. 

All of our experiments were conducted on the Radiology and the Dermatology datasets and we chose to fool four neural networks that provided the best results for the classification of these images.
In table \ref{dermatology_auc_table} we see the ROC scores for the networks EfficientNet and ResNext, on clean samples, and with the adversarial samples generated using the attacks FGSM, DeepFool, PGD, MIFGSM, and Kryptonite respectively for the dermatology (Melanoma) dataset. All of the results obtained are the average ROC scores from the five trials conducted. From table \ref{dermatology_auc_table} we can see that ResNext is \textit{more vulnerable} to adversarial samples in comparison to EffNet. We note that due to the inherent assumption of Deepfool that neural networks are all but linear it fails to outperform other iterative attacks on these datasets.

\begin{figure}[t]
\centering
\subfloat{\includegraphics[width = .16\linewidth]{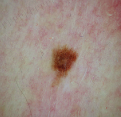}} 
\subfloat{\includegraphics[width = .16\linewidth]{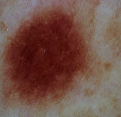}}
\subfloat{\includegraphics[width = .16\linewidth]{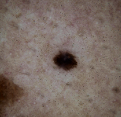}}
\hspace{1mm}
\subfloat{\includegraphics[width = .16\linewidth]{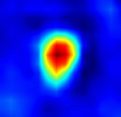}}
\subfloat{\includegraphics[width = .16\linewidth]{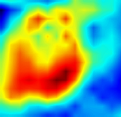}}
\subfloat{\includegraphics[width = .16\linewidth]{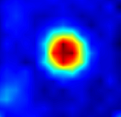}} 
\caption{Grad-CAM for Melanoma (kryptonite sample)}
\label{gradcam_melanoma}
\end{figure}

\begin{figure}[b]
\centering
\subfloat{\includegraphics[width = .16\linewidth]{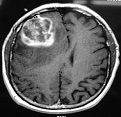}} 
\subfloat{\includegraphics[width = .16\linewidth]{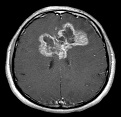}}
\subfloat{\includegraphics[width = .16\linewidth]{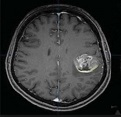}}
\hspace{1mm}
\subfloat{\includegraphics[width = .16\linewidth]{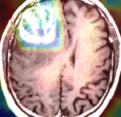}}
\subfloat{\includegraphics[width = .16\linewidth]{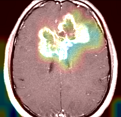}}
\subfloat{\includegraphics[width = .16\linewidth]{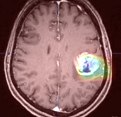}} 
\caption{Grad-CAM for MRI (kryptonite sample)}
\label{gradcam_mri}
\end{figure}

Our proposed attack as can be seen from table \ref{dermatology_auc_table} caused \textit{max drop} in network accuracy. We have utilized the Gradient-weighted Class Activation Mapping
(Grad-CAM) technique \cite{b26} which is used to find the most important regions for any given input image which affect the network output the most. From Grad-CAM image shown in figure \ref{gradcam_melanoma} we can see that the classifier takes into consideration the lesion for making its prediction. Now, since kryptonite constricts its focus as much as it can on the region of interest which is the lesion for this dataset, the classifier becomes more vulnerable when an adversarial attack targets the region on which the classification most depends on, leading to the best misclassifcation rate.

\begin{figure}[t]
\centering
\begin{tabular}{cc}
    \includegraphics[width=60mm]{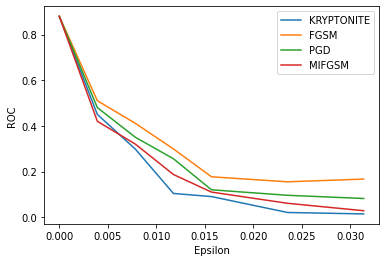} &
    \includegraphics[width=60mm]{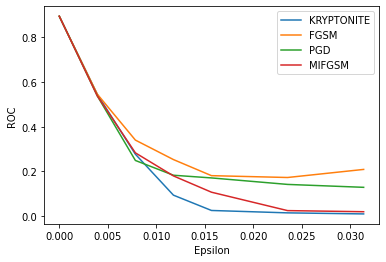} \\
    (a) EfficientNet & (b) ResNext \\
    \includegraphics[width=60mm]{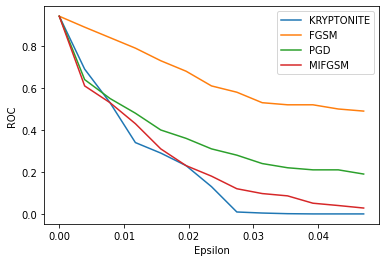} &
    \includegraphics[width=60mm]{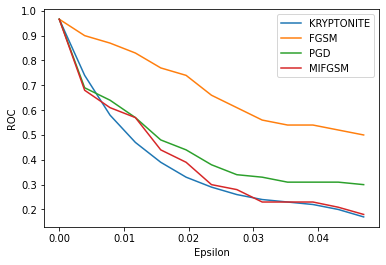} \\
    (c) Custom CNN &
    (d) VGG16 \\
\end{tabular}
\caption{Comparison of ROC score to constrain $\epsilon$}
\label{roc_vs_e_all}
\end{figure}

Furthermore, kryptonite \textit{does not} push the attention completely away from lesion in the final adversarial sample. On the contrary, the other attacks distract the classifier towards the regions completely \textit{irrelevant} for the classification which though leads to a significant drop in network accuracy but in our rigorous experiments we have seen that in kryptonite wherein all the adversarial perturbations are all but concentrated inside a fixed region (of highest interest to classifier) cause the \textit{max} misclassification rate. This is because in comparison to other adversarial samples wherein the perturbations are spread throughout the image, the classifier randomly selects the region and uses it as an important region for classification. On the other hand, a classifier fed with a kryptonite sample will always choose the lesion for classification which indeed has all of the perturbations leading to the highest drop in network accuracy.

On similar lines for the MRI dataset we can infer from table \ref{radiology_auc_table} that the Custom Convolutional Neural Network is more vulnerable to adversarial samples in comparison to VGG16 network. All of the results obtained are the average ROC scores from the six trials conducted. From the Grad-CAM images in figure \ref{gradcam_mri} we can see that for a given grayscale MRI image having a tumour, the region occupied by the tumor is most important for a classifier. In our rigorous study we found out that kryptonite restricted perturbations only to the tumour leading to malign samples being classified as benign and consequently leading to highest misclassification rates. 

\begin{figure}[t]
\centering
\begin{tabular}{cc}
    \includegraphics[width=60mm]{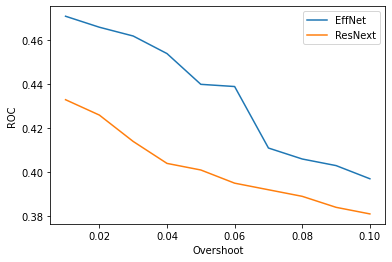} &
    \includegraphics[width=60mm]{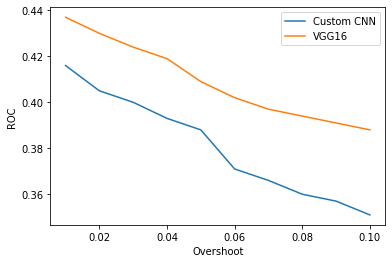} \\
    (a) Melanoma Dataset & (b) Radiology Dataset \\
\end{tabular}
\caption{Comparison of ROC score to overshoot for DeepFool}
\label{roc_df}
\end{figure}

In figure \ref{roc_vs_e_all}, it is seen that Kryptonite lowers the ROC score of all the networks tested upon better than any other attacks presented almost consistently, for perturbation sizes constrained by the hyperparameter $\epsilon$, and ultimately drops the score of generated samples to 0. Also, for DeepFool we noticed a weak negative correlation between the overshoot parameter and ROC score such that a step wise increase in that parameter led to a linear decrease in ROC score as shown in fig. \ref{roc_df}.

\begin{figure}[t]
\centering
\subfloat{\includegraphics[width = .16\linewidth]{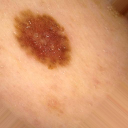}} 
\subfloat{\includegraphics[width = .16\linewidth]{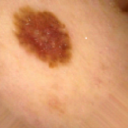}}
\subfloat{\includegraphics[width = .16\linewidth]{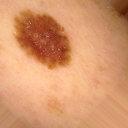}}
\subfloat{\includegraphics[width = .16\linewidth]{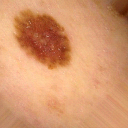}}
\subfloat{\includegraphics[width = .16\linewidth]{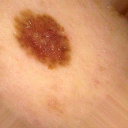}}
\subfloat{\includegraphics[width = .16\linewidth]{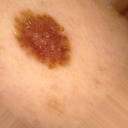}}

\subfloat{\includegraphics[width = .16\linewidth]{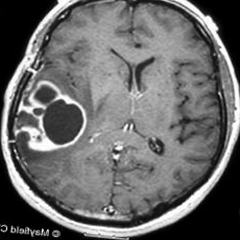}} 
\subfloat{\includegraphics[width = .16\linewidth]{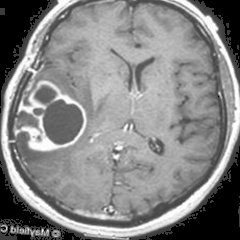}}
\subfloat{\includegraphics[width = .16\linewidth]{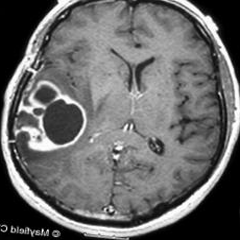}}
\subfloat{\includegraphics[width = .16\linewidth]{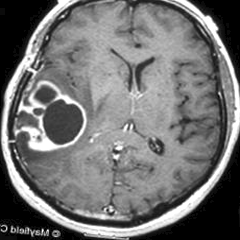}}
\subfloat{\includegraphics[width = .16\linewidth]{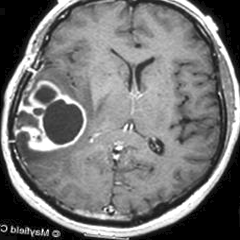}}
\subfloat{\includegraphics[width = .16\linewidth]{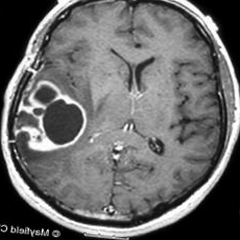}} 
\caption{Comparison of Clean, FGSM, DeepFool, MIFGSM, PGD and Kryptonite sample}
\label{adv_fig}
\end{figure}

\subsection{Perturbation size} 
Since having the highest drop in network accuracy is not the only criterion for a strong adversarial attack, as we can see from figure \ref{adv_fig} that the final kryptonite adversarial sample pertubations are almost imperceptible and looks very similar to the clean sample to a naked eye. 

In the tables \ref{dermatology_auc_table} and \ref{radiology_auc_table}, we can observe the average and worse case perturbation size for each of the attack and the corresponding network accuracy (at average perturbation level). Furthermore, we note that DeepFool provides the minimal perturbations as it was designed to do so but kryptonite is a powerful alternative as well with highly respectable perturbation sizes in comparison to DeepFool and \textit{highest} drop in network accuracy.
As mentioned before kryptonite ensures the classifier uses the lesion/tumor for classification, consequently, \textit{smaller} perturbations to the same cause high misclassification rates as can be seen from figure \ref{roc_vs_e_all}.

\begin{table*}[t]
\tiny
\caption{The table below shows the average (5 trials conducted) of adversarial defence results for the proposed attack. The metrics on the left of the slash represent accuracy of the model with adversarial samples as input. The metrics on the right of the slash represents the accuracy on clean samples.}
\label{defence_table}
\centering
\begin{threeparttable}
\begin{tabular}{ |c|c|c|c|c|c|c|c|c|c|c| } 
\hline
\multirow{2}{*}{Melanoma} & \multicolumn{2}{|c|}{FGSM} & \multicolumn{2}{|c|}{DeepFool} 
& \multicolumn{2}{|c|}{PGD } & 
 \multicolumn{2}{|c|}{MIFGSM} & \multicolumn{2}{|c|}{Our's}\\
\cline{2-11}
& EffNet & ResNext & EffNet & ResNext & EffNet & ResNext & EffNet & ResNext &  EffNet & ResNext\\
\hline
\shortstack{Adv.Train \\(FGSM)} & 0.71/0.82 & 0.79/0.86 & 0.48/0.82 & 0.54/0.86 & 0.29/0.82 & 0.33/0.86 & 0.28/0.82 & 0.29/0.86 & 0.19/0.82 & 0.27/0.86 \\
\hline
\shortstack{Adv.Train \\(DeepFool)} & 0.69/0.84 & 0.73/0.85 & 0.78/0.84 & 0.80/0.85 & 0.46/0.84 & 0.47/0.85 & 0.40/0.84 & 0.44/0.85 & 0.31/0.84 & 0.32/0.85 \\
\hline
\shortstack{Adv.Train \\(PGD)} & \textbf{0.73}/0.76 & \textbf{0.75}/0.78 & \textbf{0.73}/0.76 & \textbf{0.74}/0.78 & 0.67/0.76 & 0.70/0.78 & 0.64/0.76 & \textbf{0.69}/0.78 & 0.58/0.76 & 0.60/0.78 \\
\hline
\shortstack{Adv.Train \\(MIFGSM)} & \textbf{0.62}/0.65 & \textbf{0.66}/0.68 & \textbf{0.59}/0.65 & \textbf{0.60}/0.68 & \textbf{0.58}/0.65 & 0.58/0.68 & 0.51/0.65 & 0.55/0.68 & 0.47/0.65 & 0.52/0.68 \\
\hline
\shortstack{Adv.Train \\(Our's)} & \textbf{0.69}/0.72 & \textbf{0.70}/0.74 & \textbf{0.65}/0.72 & \textbf{0.66}/0.74 & 0.62/0.72 & \textbf{0.65}/0.74 & 0.59/0.72 & \textbf{0.65}/0.74 & 0.57/0.72 & 0.62/0.74 \\
\hline
PD & \textbf{0.49}/0.56 & \textbf{0.50}/0.59 & 0.44/0.56 & 0.48/0.59 & 0.39/0.56 & 0.41/0.59 & 0.30/0.56 & 0.33/0.59 & 0.20/0.56 & 0.23/0.59 \\
\hline
DD & \textbf{0.85}/0.86 & \textbf{0.85}/0.88 & \textbf{0.83}/0.86 & \textbf{0.84}/0.88 & \textbf{0.77}/0.86 & \textbf{0.80}/0.88 & 0.74/0.86 & \textbf{0.79}/0.88 & 0.70/0.86 & 0.73/0.88\\
\hline
\multirow{2}{*}{MRI} & \multicolumn{2}{|c|}{FGSM} & \multicolumn{2}{|c|}{DeepFool} 
& \multicolumn{2}{|c|}{PGD } & 
 \multicolumn{2}{|c|}{MIFGSM} & \multicolumn{2}{|c|}{Our's}\\
\cline{2-11}
& CNN & VGG16 & CNN & VGG16 & CNN & VGG16 & CNN & VGG16 & CNN & VGG16\\
\hline
\shortstack{Adv.Train \\(FGSM)} & 0.81/0.88 & 0.84/0.93 & 0.42/0.88 & 0.47/0.93 & 0.38/0.88 & 0.41/0.93 & 0.24/0.88 & 0.30/0.93 & 0.18/0.88 & 0.23/0.93 \\
\hline
\shortstack{Adv.Train \\(DeepFool)} & 0.79/0.90 & \textbf{0.86}/0.91 & 0.76/0.90 & 0.78/0.91 & 0.43/0.90 & 0.50/0.91 & 0.25/0.90 & 0.33/0.91 & 0.17/0.90 & 0.25/0.91 \\
\hline
\shortstack{Adv.Train \\(PGD)} & \textbf{0.80}/0.83 & \textbf{0.82}/0.84 & \textbf{0.78}/0.83 & \textbf{0.79}/0.84 & 0.71/0.83 & 0.75/0.84 & 0.66/0.83 & 0.70/0.84 & 0.57/0.83 & 0.63/0.84 \\
\hline
\shortstack{Adv.Train \\(MIFGSM)} & \textbf{0.74}/0.76 & \textbf{0.75}/0.79 & \textbf{0.67}/0.76 & 0.69/0.79 & 0.56/0.76 & 0.62/0.79 & 0.69/0.76 & 0.73/0.79 & 0.44/0.76 & 0.49/0.79 \\
\hline
\shortstack{Adv.Train \\(Our's)} & \textbf{0.80}/0.82 & \textbf{0.81}/0.85 & \textbf{0.76}/0.82 & \textbf{0.78}/0.85 & \textbf{0.73}/0.82 & \textbf{0.76}/0.85 & 0.68/0.82 & 0.69/0.85 & 0.64/0.82 & 0.69/0.85 \\
\hline
PD & \textbf{0.51}/0.60 & \textbf{0.60}/0.67 & \textbf{0.53}/0.60 & \textbf{0.63}/0.67 & 0.47/0.60 & 0.53/0.67 & 0.46/0.60 & 0.51/0.67 & 0.16/0.60 & 0.19/0.67 \\
\hline
DD & \textbf{0.92}/0.93 & \textbf{0.94}/0.96 & \textbf{0.89}/0.93 & \textbf{0.93}/0.96 & \textbf{0.86}/0.93 & \textbf{0.89}/0.96 & \textbf{0.84}/0.93 & 0.85/0.96 & 0.79/0.93 & 0.84/0.96\\
\hline
\end{tabular}
\begin{tablenotes}
      \tiny
      \item In the table above values mentioned in bold font indicate resilience to a particular attack. We have chosen a drop of less than 10\% in accuracy from the original accuracy of a particular network on clean samples as a sign of robustness or resilience to a particular attack. 
      \item For example, consider PGD Adversarial Training for Melanoma dataset performed on ResNext network. Now attacking this network with MIFGSM adversarial samples results in an accuracy of 69\% which is a less than 10\% drop from the accuracy on clean samples (78\%). Hence, we say that performing PGD adversarial training on ResNext makes it robust to MIFGSM.
      \item If an adversarial training defence trained on a particular attack is implemented on a network to counter the same attack, the effectiveness of the implemented defence does not indicate any kind of robustness.
    \end{tablenotes}

\end{threeparttable}
\end{table*}

\subsection{Adversarial Defence for Kryptonite}
We have evaluated our proposed method on three types of adversarial defence, the detailed results and analysis is provided in table \ref{defence_table}. These include Adversarial Training \cite{b3}, Pixel Deflection (PD)\cite{b27} and Defensive Distillation (DD) \cite{b28}. The results are an average of five rigorous trials conducted. The \textit{JumpReLU} defence \cite{b13} was easily broken with increasing perturbation size, hence, was avoided in this study.

\subsubsection{Adversarial Training} We performed adversarial training using each of the attack algorithms using both of our datasets on their respective classifiers. To ensure the classifier does not fail to classify on clean samples we trained the classifiers on complete training data which included 65\% adversarial samples and 35\% clean samples. All the adversarial attacks were kept at their optimal parameters while evaluating defence as well. As we can see from table \ref{defence_table} re - training classifiers using FGSM and DeepFool did not help much in making them more robust against adversaries. On the other hand, PGD adversarial training which is considered the best was able to provide robustness against FGSM, DeepFool and even MIFGSM (for ResNext). We noticed that PGD adversarial training was really slow as compared to others. Adversarial training done using MIFGSM helped make the classifiers robust against FGSM, DeepFool and PGD (for EffNet). Furthermore, adversarial training performed using the samples generated by our proposed method show that we were able to make our classifiers robust to FGSM, DeepFool, PGD (excluding EffNet case) and even MIFGSM (for ResNext). In terms of accuracy on clean images, a classifier trained using DeepFool samples performs the best owing to its small perturbation size but this does not ensure robustness whereas using PGD or Kryptonite adversarial training though we lose out a bit on accuracy on clean images we still have a higher chance towards resilience to adversaries. Adversarial training gives a \textit{false} sense of security \cite{b29} and for very large scale datasets it seems impractical to re - train the huge neural nets just for certain amount of robustness. Furthermore, the robustness achieved through this method seems limited as we can infer from table \ref{defence_table} that this method could not bring robustness towards our proposed attack. Another observation made was that going beyond optimal parameters resulted in breaking this defence at the cost of perceptible perturbations. 

\subsubsection{Pixel Deflection} In this method proposed by \cite{b27}, we randomly select a pixel and replace that pixel with another randomly selected pixel, this process the authors have called as pixel deflection (PD). Then, we use a Robust Class Activation Map (RCAM) to select pixel which is least important for classification and deflect that pixel. The added noise is then removed using Wavelet Denoising (WD). The authors of this defence base their intuition on the fact that adversarial attacks are \textit{antagonistic} to region's of interest but for kryptonite region of interest is everything. Hence, as we can see from table \ref{defence_table}, this defence \textit{struggles} to provide any kind of resilience towards our proposed attack. Another observation was that this method led to a big drop of network accuracy on clean images. The parameter deflections was set at 120, 100, 100, 80 for the networks EffNet, ResNext, CNN and VGG16 respectively. Pixel Deflection performs the best in terms of computational efficiency but with increasing $\epsilon$ this approach failed as well. Also, we found out that in some cases the amount of noise even after performing Wavelet Denoising was really high which lead to poor classification performance on clean images as information critical to the classifier got deflected as well. 

\subsubsection{Defensive Distillation} Proposed by \cite{b28}, the idea is to train a neural network having a softmax output layer at some temperature \(T\) using the hard or discrete labels. Then, the obtained class probabilities or soft labels are used to train another neural network with the same architecture as the original one which is called a distilled network. In our experiments the parameter \(T\) was set at 10, 15, 20 and 20 for the classifiers EffNet, ResNext, CNN and VGG16 respectively. From table \ref{defence_table} we can see that this defence provides very good resilience to all of the attacks at the same time also retains accuracy on clean samples but this idea just makes the \textit{generation} of adversarial samples difficult. Secondly, it has been shown by Carlini in \cite{b30} that distillation can be easily broken. In our preliminary testing using the method proposed by \cite{b30}, we saw a misclassification rate of a respectable 64\% for our proposed method. Also, increasing perturbation size to perceptible levels led to an increased misclassification rate for this method of defence. Defensive distillation hides the gradient, making it tough for us to find adversarial samples. We found out that this method flattens out the model completely in comparison to adversarial training. Furthermore, in our preliminary black - box testing wherein we attacked a distilled model using adversarial samples generated by another model having similar architecture, our attack produced a misclassification rate of 79\% (conducted on the MRI dataset). Furthermore, we plan to implement more advanced defence and adversarial sample detection techniques for this proposed attack.

\subsection{Time Efficiency Comparison}
Table \ref{efficiency_table} shows the efficiency of our proposed attack. Though FGSM takes the least amount of time to generate an adversarial sample, its a comparatively weak attack as shown before. Furthermore, we are performing better in terms of efficiency in comparison to PGD and DeepFool attacks. These results further extend towards adversarial training as well with PGD and DeepFool being the slowest, FGSM the fastest while, MIFGSM and our proposed method perform similarly.

\begin{table}[t]
\caption{The table below shows the average of time (10 trials conducted) required to generate a single adversarial sample (in seconds) with all the hyper parameters of the attacks at their optimal values.}
\centering
\begin{tabular}{| c | c | c | c | c |}
\hline
\multirow{2}{*}{Attack} & \multicolumn{2}{|c|}{Melanoma} & \multicolumn{2}{|c|}{MRI} \\
\cline{2-5}
\label{efficiency_table}
& EffNet & ResNext & CNN & VGG16\\ 
\hline
FGSM & 0.23 & 0.19 & 0.11 & 0.14 \\
\hline
DeepFool & 0.96 & 0.90 & 0.84 & 0.87 \\
\hline
PGD & 0.75 & 0.70 & 0.61 & 0.64 \\
\hline
MIFGSM & 0.43 & 0.41 & 0.32 & 0.35 \\
\hline
KRYPTONITE & 0.44 & 0.41 & 0.31 & 0.35\\
\hline
\end{tabular}
\end{table}

\section{Limitations of the study and Future work}\label{limitations}
The attack presented in this study has only been evaluated upon certain medical datasets due to difficulties in generalising the attack. A limitation of our attack is that identifying a region of interest for more elaborate images is challenging. In the future, we hope to advance the scope of this algorithm to identify regions of interest for diverse images and hence present results for standard datasets.
We also hope to motivate the creation of a more robust defence in order to protect the main region of interest.

\section{Conclusion}\label{conclusion}
In this paper we have presented the adversarial attack Kryptonite, which is used to find the most optimal perturbation that can be added to an image in order to encourage a network to make an incorrect prediction. This attack that hides in plain sight, uses the changes in a region of interest to find the least possible added noise in a constricted region to fool a given network. 
The effectiveness of this attack, as seen with the given empirical evidence, shows that the region of interest of an image is a major vulnerability in an image that can be exploited. Furthermore, it is seen that more changes constricted to this region can mitigate the effectiveness of some state of the art defences, as compared to those attacks that are more antagonistic to the region of interest. We have hence shown that localising the added perturbation makes an attack more robust.

\end{document}